\documentclass[letterpaper, 10 pt, conference]{ieeeconf}  
\IEEEoverridecommandlockouts
\usepackage{comment}
\usepackage{cite}
\usepackage{amsmath,amssymb,amsfonts}
\usepackage{amsthm}
\usepackage[ruled,vlined]{algorithm2e}
\usepackage{gensymb}
\usepackage{caption}
\usepackage{siunitx}
\usepackage{graphicx}
\usepackage{booktabs}
\usepackage{textcomp}
\usepackage{mathtools}
\usepackage[caption=false]{subfig}
\usepackage{xcolor}
\usepackage{hyperref}
\usepackage{todonotes}
\usepackage{amsfonts}
\usepackage{amssymb}
\usepackage{amsmath}
\usepackage{hyperref}
\usepackage{xcolor}
\usepackage{user_macros}
\usepackage{soul}
\usepackage{diagbox}
\usepackage[nolist,nohyperlinks]{acronym}
\def\BibTeX{{\rm B\kern-.05em{\sc i\kern-.025em b}\kern-.08em
    T\kern-.1667em\lower.7ex\hbox{E}\kern-.125emX}}
    
\newcommand{\sgnote}[1]%
{\textcolor{green}{\textbf{Note: #1}}}
\newcommand{\shnote}[1]%
{\textcolor{blue}{\textbf{Note: #1}}}

\usepackage{siunitx} 

\newcommand{\nhnote}[1]{\textcolor{blue}{NH: #1}}

\begin{document}
\bstctlcite{IEEEexample:BSTcontrol}
\begin{acronym}
\acro{HJ}{Hamilton-Jacobi}
\acro{HJI}{Hamilton-Jacobi-Isaacs}
\acro{DP}{Dynamic Programming}
\acro{ODE}{Ordinary Differential Equation}
\acro{MPC}{Model Predictive Control}
\acro{MDP}{Markov Decision Processes}
\acro{RBF-NN}{Radial Basis Function Neural Network}
\acro{RMSE}{Root Mean Squared Error}
\acro{MSE}[MSE]{mean-squared-error}
\acro{RL}{Reinforcement Learning}
\acro{PDE}{Partial Differential Equation}
\acro{ASV}{Autonomous Surface Vehicle}
\acro{MAS}{Multi-Agent-Systems}
\acro{H-MAS}{Hierarchical Control of Multi-Agent-Systems}
\acro{IPM}{Isolated Platform Metric}
\acro{FRS}{Forward Reachable Set}
\acro{IPM}{Isolated Platform Metric}
\acro{HJR}{Hamilton Jacobi Reachability}
\acro{mt-HJR}{multi-time Hamilton Jacobi Reachability}
\acro{mt-HJR-B}{multi-time Hamilton Jacobi Reachability Baseline}
\acro{mt-HJR-RC}{multi-time Hamilton Jacobi Reachability with multi-agent Reactive Control}
\acro{mt-HJR-LISIC}{multi-time Hamilton Jacobi Reachability with multi-agent LISIC}
\acro{HJR-Reactive}{Hamilton Jacobi Reachability Reactive Control}
\acro{HJR-Flocking}{Hamilton Jacobi Reachability Flocking Control}
\acro{CBFs}{Control Barrier Functions}
\acro{LISIC}{Low Interference Safe Interaction Controller}
\acro{CDR}{Carbon Dioxide Removal}
\end{acronym}

\title{\LARGE \bf
Safe Connectivity Maintenance of Underactuated Multi-Agent Networks in Dynamic Oceanic Environments}

\author{Nicolas Hoischen$^{1,2,*}$, Marius Wiggert$^{2,*}$,
and Claire J. Tomlin$^{2}$
\thanks{$^{*}$ Both authors contributed equally to this research. $^{1}$ N.H. is with the School of Computation, Information and
Technology of the Technical University of Munich, Germany, $^{2}$ N.H., M.W., and C.J.T. are with EECS at the University of California, Berkeley, USA. For inquiries contact: {\tt\small mariuswiggert@berkeley.edu}}
\thanks{The authors gratefully acknowledge the C3.ai Digital Transformation Institute, NASA ULI on Safe Aviation Autonomy, DARPA Assured Autonomy Program, and the Zeno Karl Schindler Foundation.}
}

\maketitle

\begin{abstract}
Autonomous multi-agent systems are increasingly being deployed in environments where winds and ocean currents have a significant influence. Recent work has developed control policies for single agents that leverage flows to achieve their objectives in dynamic environments. However, in multi-agent systems, these flows can cause agents to collide or drift apart and lose direct inter-agent communications, especially when agents have low propulsion capabilities. To address these challenges, we propose a hierarchical multi-agent control approach that allows arbitrary single-agent performance policies that are unaware of other agents to be used in multi-agent systems while ensuring safe operation.
We first develop a safety controller using potential functions, solely dedicated to avoiding collisions and maintaining inter-agent communication. 
Next, we design a low-interference safe interaction (LISIC) policy that trades off the performance policy and the safety control to ensure safe and performant operation. Specifically, when the agents are at an appropriate distance, LISIC prioritizes the performance policy while smoothly increasing the safety controller when necessary. We prove that under mild assumptions on the flows experienced by the agents, our approach can guarantee safety. Additionally, we demonstrate the effectiveness of our method in realistic settings through an extensive empirical analysis with simulations of fleets of underactuated autonomous surface vehicles operating in dynamic ocean currents where these assumptions do not always hold.

\end{abstract}

%

\section{Introduction}
\label{sec:introduction}

\begin{figure}[t]
\vspace{0pt}
\includegraphics[width=0.5\textwidth, trim={2cm 0cm 6cm 0cm}, clip]{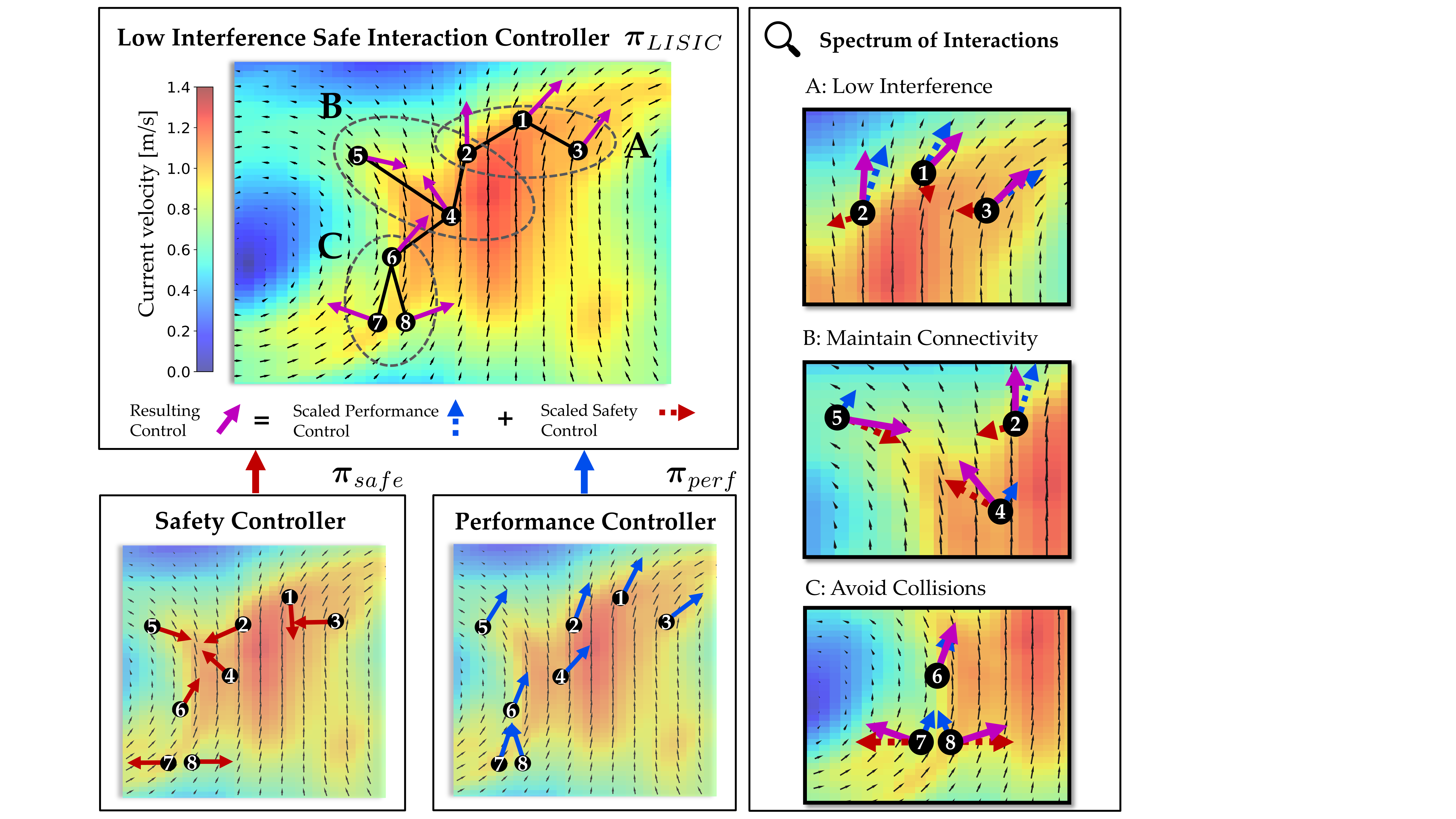}
\caption{\small 
Our \ac{LISIC} policy blends a single-agent performance control input with a flocking-based safety control input to avoid connectivity losses and collisions in a multi-agent network while minimally interfering with the performance objective of each agent. This ensures safe performance in ocean environments with strong ocean currents affecting the low-powered agents.}
\label{fig:head_picture}
\vspace{0pt}
\end{figure}

Autonomous multi-agent systems, from drones to balloons and ocean surface vessels, are increasingly being explored for various applications, including inspection, collecting data, or scaling ocean aquaculture \cite{sujit2009uav, phykos}. In many applications, the agents communicate with each other for various purposes: to achieve a joint objective, to ensure internet coverage \cite{project_loon}, or to share information amongst each other to improve operations. 
Local communication often relies on limited-range systems, e.g., sonar or radar, requiring agents to stay close to each other for network connectivity (see Fig. \ref{fig:head_picture}). 

When a robotic system operates in the oceans and air, it is exposed to winds and currents. Most control approaches consider these as disturbances for which an overactuated control needs to compensate. What if, instead, the agent takes advantage of these flows? Recent work demonstrated that by \textit{going with the flow} and using small actuation strategically to nudge itself into favorable flows, an agent can achieve its objective with very little energy \cite{
wiggert_etal_CDC2022, killer_CDC2023, bellemare2020autonomous, doshi_etal_CDC2022, doering_CDC2023}.

Given such individual agent performance controllers \cite{wiggert_etal_CDC2022}, we aim to develop a method that extends to multi-agent systems operating in complex flows while ensuring network connectivity and avoiding collision among agents. From the control perspective, this is challenging because of two key reasons. First, disconnections are sometimes unavoidable in the underactuated setting, where the agent's individual propulsion is smaller than the surrounding flows, as the nonlinear, time-varying flows can push agents in opposing directions. The safe interaction controller needs to be resilient and recover connectivity after losing it. Second, constraint satisfaction needs to be traded off intelligently with the performance objective of each agent. For example, a time-optimal controller for an agent would prefer staying in strong flows, which can conflict with the network connectivity objective. Our insight is that we can simplify this multi-agent problem using three different controllers in a \ac{H-MAS} approach (Fig. \ref{fig:head_picture}). \\

\textit{Related Literature.} In \ac{H-MAS}, agents are organized into multiple levels of hierarchy, with higher-level agents having more authority and control over lower-level agents, designated as followers \cite{douMultiAgentBasedHierarchical2013}. For instance, \cite{lollaPathPlanningMultiscale2015} solves path planning and ocean vehicle coordination separately with a leader-follower structure. For distance-based control tasks, such as for the Safety Controller in Fig. \ref{fig:head_picture}, flocking techniques can maintain connectivity by influencing the agent's behavior to follow the movement of their neighbors while avoiding collisions \cite{zavlanosGraphtheoreticConnectivityControl2011a, olfati-saberFlockingMultiAgentDynamic2006}. Recent advancements in \ac{MPC} have also achieved connectivity and collision-free operation within \ac{MAS} \cite{carron2023multi, filotheouDecentralizedControlUncertain2018} and successfully approached control of varying-topology networks \cite{PnP}. Nevertheless, a notable limitation of these approaches is their reliance on the assumptions of position invariance \cite{carron2023multi}, fixed neighbor sets \cite{filotheouDecentralizedControlUncertain2018} and time invariance \cite{carron2023multi, PnP, filotheouDecentralizedControlUncertain2018} of the flows, which do not apply in dynamic ocean environments. 
Thus far, the mentioned literature offers limited applicability to time-varying, uncertain flows predicted by forecasts. Since the agents are often underactuated, they cannot reliably compensate for disturbances. In this context, optimization-based approaches such as \ac{MPC} frequently lead to increased computational complexity, convergence to local minima, and infeasibility with respect to the constraints. Hence, we opt for computationally efficient flocking approaches to achieve distance-based safety control, which are always feasible and, when coupled with effective single-agent planners \cite{wiggert_etal_CDC2022}, can be run at high update rates.
While many flocking schemes only assume simple double integrator dynamics, adaptive flocking has also been applied to nonlinear dynamics \cite{yuRobustAdaptiveFlocking2010a, zhangAdaptiveFlockingNonlinear2015, zouFlockingUncertainNonlinear2021, pengNeuralAdaptiveFlocking2012}. However, due to the time-varying dynamics, adaptive approaches may not generalize well, especially given the diverse nature of currents. 

\textit{Contributions.} To address the above shortcomings, we propose a \acl{LISIC} (Fig. \ref{fig:head_picture}). This framework is more general in three dimensions. First, it takes an arbitrary performance policy, in contrast to \cite{olfati-saberFlockingMultiAgentDynamic2006, suFlockingMultiAgentSystems2012, liFlockingMultiAgentSystems2016}, where the flock can only track reference trajectories of single or multiple virtual leaders. In fact, a feedback control policy can optimize objectives besides navigation and, in complex flows, leads to significantly better results than tracking a reference trajectory \cite{wiggert_etal_CDC2022}. This enables the use of \ac{DP} approaches where the value function yields optimal individual agent controls for an arbitrarily high number of agents without additional cost beyond a cheap gradient computation. This is especially powerful for multi-agent problems where the objective can be decomposed into the sum of independent single-agent objectives. Second, we provide design choices to modulate the aggressiveness of pursuing safety versus performance. Third, our approach also enables recoveries in case of connectivity losses. Finally, we investigate our method in the context of a promising approach to \acf{CDR}: utilizing robotic seaweed farms \cite{phykos}.

\textit{Organization.} Section \ref{sec:ProblemFormulation} introduces relevant background and metrics to evaluate our \ac{LISIC} in complex flows. In Section \ref{sec:method}, we present our \ac{LISIC} approach, whereas in Section \ref{sec:theoreticalAnalysis}, we prove that our method guarantees safe network interactions under certain conditions on the maximum magnitude of the control and flow field velocities across the agents. Finally, we assess the performance of our approach in realistic ocean currents where these conditions are not always met with the metrics defined in Section\ref{sec:ProblemFormulation}.

\section{Problem Formulation}
\label{sec:ProblemFormulation}
In this section, we first describe the system's dynamics and briefly summarize connectedness in communication graphs. Then, we define our problem statement and the metrics we use to measure constraint violation.

\subsection{System Dynamics}
We consider a swarm of $N$ agents and use $\mathcal{V}$ to describe the set of all agents. Let the actuation signal of each agent $i$ be denoted by $\ui$ from a bounded set $\mathbb{U}\in \R^{n_u}$ where $n_u$ is the dimensionality of the control.
Then, the dynamics for each agent $i \in \mathcal{V}$ are given by:
\begin{equation}
        \dot{\qi} = v(\qi,t) + g(\qi,\ui, t), \; t \in [0, T]
    \label{eq:system_dynamics}
\end{equation}
$\qi \in \R^n$ denotes the position of agent $i$ in the $n$ dimensional state space, where $n=2$ for a surface vessel on the ocean. The movement of agent $i$ depends on the time-varying non-linear flow field $v(\qi,t) \to \R^n$ and its control $g(\qi,\ui, t)$. Although our method works for arbitrary $g$, we will focus on situations where the agent can directly actuate in each dimension, i.e., $g(\qi,\ui, t) = \ui$, in line with our experiments discussed in Section \ref{sec:simulation}.
Let the agent trajectory resulting from Eq. (\ref{eq:system_dynamics}) be described by $\traj_i$ with $\traj_i(t)$ the state $\qi$ at $t$. For the global system of all $N$ agents, we use $\q = [\q_1^\top, \q_2^\top, \ldots,\q_N^\top]^\top$, $\bu = [\bu_1^\top, \bu_2^\top, \ldots, \bu_N^\top]^\top$, and $\traj$ respectively to describe the state, control, and trajectory. 

\begin{remark}
    While our method also works in known currents, we focus on realistic settings, where only coarse forecasts $\hat{v}$ are available to the planner, which differ significantly from the true flows $v$.
\end{remark}

\subsection{Communication Graph Preliminaries}
\label{subsec:graph_theory}

The network topology of our \ac{MAS} with state $\q$ can be represented by an undirected communication graph $G(t)$, allowing information to flow bidirectionally between agents. The set of finite vertices $\mathcal{V} = \{1, 2 \dotsc N\}$ denotes individual agents, while the time-varying set of edges $\mathcal{E}(t) \subseteq \left\{(i,j) \in (\mathcal{V} \times\mathcal{V}), j \neq i\right\}$ represents direct communication between agents. Given $\rcom$ an upper communication threshold, $(i,j) \in \mathcal{E}(t) \iff  d(\qi, \qj) < \rcom$. In other words, agents $i$ and $j$ can communicate directly with each other if they are spatially close with respect to a distance measure $d(\qi, \qj)$. The graph $G(t)$ is said to be \emph{connected} if an undirected path exists between every pair of distinct vertices. 
Next, we define the the $N \times N$ adjacency matrix $A(G(t)) = \left[a_{ij}(t)\right] \in \{0,1\}$ encoding the connectivity between vertices, i.e. $a_{ij}(t) = 1 \iff (i,j) \in \mathcal{E}(t)$.

The degree of a vertex $i$ at time $t$, $deg(i, t) = \sum_{j=1}^N A_{i j}(t)$ represents the number of incident edges to vertex $i$. 
The degree matrix $D$ is then defined as the diagonal matrix $D(G(t)) =\operatorname{diag}\left(deg(i, t)\right)$. To measure the graph's connectivity, we can compute the eigenvalues of the Laplacian positive semi-definite matrix $L= D(G(t)) - A(G(t))$. The second smallest eigenvalue $\lambda_2(L(G(t))$, commonly referred to as the algebraic connectivity or Fiedler value, captures the robustness of the network to link failures. In particular, $G(t)$ is connected if and only $\lambda_2(L(G(t))) > 0$  \cite{zavlanosGraphtheoreticConnectivityControl2011a}.

\subsection{Problem Statement}
\label{sec:ProblemStatement}

We focus on multi-agent problems where the joint objective is the sum of independent objectives $\mathbb{P}_i$, which can be sketched out as:
\begin{subequations} \label{eq:pb_statement}
\begin{align}
  \min_{\pi} \quad & \sum_{i=1}^N \mathbb{P}_i(\traj_i, \ui(\cdot)) \\
  \textrm{s.t.} \quad & \forall \; t \in [t_0, T] \nonumber\\
  & \dot{\traj}(t) = v(\traj(t),t) + \pi(\traj(t)) \quad \text{global dynamics (\ref{eq:system_dynamics})} 
  \nonumber
  \\
    & d(\traj_i(t), \traj_j(t)) > \rcoll \quad (i,j) \in \mathcal{V} \times \mathcal{V}, i \neq j \label{eq:coll_av}\\
    & \lambda_2(L(G(\traj(t), \rcom))) > 0 \label{eq:conn_constraint}
\end{align}
\end{subequations}
We aim to find a control policy $\pi$ that approximately solves Eq. (\ref{eq:pb_statement}) while being computationally cheap and always feasible. The agents are coupled in only two constraints:  the collision constraint (\ref{eq:coll_av}) where $d(\qi, \qj)$ represents the distance between agent $i$ and $j$ and $\rcoll$ the minimum safe distance, and second, Eq. (\ref{eq:conn_constraint}), in maintaining a graph where all agents are connected based on the communication range $\rcom$. 
\begin{assumption}
    In real-world scenarios,  the initial network may begin in a disconnected state or transition to a disconnected state when underactuated agents are involved. Hence, it is assumed that each agent possesses an \textit{emergency communication backup} to a central unit, e.g., via satellite, in case its closest distance to any other agent exceeds $R_{com}$. The objective is to minimize such instances, as these forms of communication are typically expensive.
\end{assumption}

Our insight is that in this setting, we can decompose the problem and handle the objectives and constraints on different levels with (1) a performance controller $\piperf$ for each agent, (2) a safety controller $\pisafe$, and (3) a low-interference safe interaction controller $\pilisic$ trading-off the two (Fig. \ref{fig:head_picture}). 

The performance controller of an agent $i$ minimizes its $(\piperf)_i = \argmin_{\pi_i} \mathbb{P}_i(\traj_i, \ui(\cdot))$ only considering its own dynamics (\ref{eq:system_dynamics}). $\piperf$ can be an arbitrary control policy from a fixed control signal to a feedback controller based on learning or dynamic programming (Section \ref{sec:simulation}). In challenging settings like ours with non-linear, time-varying dynamics, it is easier to design single-agent feedback controllers than solving the coupled multi-agent problem above, e.g., for time-optimal navigation, reference tracking, or optimizing seaweed growth \cite{killer_CDC2023}. The safety controller, $\pisafe$, determines the control for all agents to ensure the interaction constraints (\ref{eq:coll_av}), (\ref{eq:conn_constraint}) are satisfied. Lastly, based on the control inputs $\uperf$ and $\usafe$ from the respective policies, the safe interaction controller 
decides the agents final control inputs $\bu = \pilisic(\uperf, \usafe)$. To achieve good performance, the safe interaction controller should not interfere too much with $\uperf$ while still ensuring connectivity and avoiding collisions. This work focuses on designing $\pisafe$ and $\pilisic$ for an arbitrary $\piperf$.

\subsection{Evaluation Metrics} 
\label{subsect:metrics}

Due to the underactuated nature of the agents, it is impossible to guarantee network connectivity or collision avoidance in some scenarios. Hence, we define evaluation metrics to assess the safety of our control schema with respect to Eq. (\ref{eq:conn_constraint}) and (\ref{eq:coll_av}). A collision happens between any of the agents in the swarm if $\exists t \in [0,T]$ at which Eq. (\ref{eq:coll_av}) is violated. We denote this with the collision indicator $\mathbb{I}_{coll} \in \{0,1\}$. To measure various aspects of losing connectivity, we use three metrics. First, for a binary measure, if disconnections occur, we define the disconnection indicator $\mathbb{I}_{disconn} \in \{0,1\}$ which is $1$ if $\exists t \in [0,T]$ at which Eq. (\ref{eq:conn_constraint}) is violated and zero otherwise. 
Additionally, we measure the minimum Fiedler value over time; the higher, the more robust the communication network (Section \ref{subsec:graph_theory}):
\begin{equation}
    \lambda_2^{min} = \min_{t \in [0, T]} \lambda_2(L(G(\traj(t), \rcom))
    \label{eq:fiedler}
\end{equation}

Lastly, as single-agent backup communication is costly, it matters how long an agent is isolated from all other agents. Therefore, we are introducing a new measure called \ac{IPM}.
\begin{equation}
    \textsc{IPM} = \frac{1}{T}\int_{t=0}^T M(deg(i, t)=0)\; dt
    \label{eq:IPM}
\end{equation}
where $M(deg(i, t)=0)$ counts the number of disconnected vertices, which corresponds to the number of zeros in the diagonal of the graph degree matrix $D(G(\traj(t),\rcom)$ (Section  \ref{subsec:graph_theory}).

In Section \ref{sec:simulation}, we compare different controllers empirically over a large, representative set of missions $\mathbb{M}$ by evaluating the collision rate $\mathbb{E}_{\q(t_0), t_0 \sim \mathbb{M}}\left[\mathbb{I}_{coll}\right]$, the disconnection rate $\mathbb{E}_{\q(t_0), t_0 \sim \mathbb{M}}\left[\mathbb{I}_{disconn}\right]$, as well as the distributions of IPM and $\lambda_2^{min}$. In our setting where the performance objectives $\mathbb{P}_i$ are minimum time-to-target for each agent $i$, the connectivity constraint often leads to a trade-off with the performance objective. Hence, we also quantify the degradation of the performance controller by quantifying the minimum distance the swarm center got to the target area $\mathcal{T}$ over the mission time $t \in [0, T]$ as $d_{min}(\mathcal{T})$.

\section{Method}
\label{sec:method}

Our method tackles the multi-agent problem with a hierarchical control approach. The low interference safe interaction controller $\pilisic$ ensures performance and safe control based on an arbitrary performance controller $\piperf$ and a safety controller $\pisafe$ (see Fig. \ref{fig:head_picture}). As explained in Section \ref{sec:introduction}, our approach is more general than \cite{suFlockingMultiagentSystems2008, suFlockingMultiAgentSystems2012, liFlockingMultiAgentSystems2016}, as an arbitrary performance policy is chosen and our framework allows for designing the aggressiveness of the safety flocking-based controller following the magnitude of the performance control input. We first introduce our flocking-inspired safety controller based on potential functions and then detail our design for $\pilisic$. Note that this control scheme is applicable in fully actuated agents, even though our primary focus lies in applications involving underactuated agents, which initially motivated a feasible reactive approach to collision avoidance and connectivity.

\subsection{Flocking-Inspired Safety Controller}
\label{subsec:safetyctrl}


The sole objective of the safety controller is to ensure adequate distances between the agents without prescribing a formation. Hence, we design our safety controller $\pisafe$ based on the gradients of a potential function $\psi$. 

To explain the principle, let us first focus on two connected agents $i$ and $j$ at an inter-agent distance $\norm{\qij} = \norm{\qi - \qj}$. Consider the following bowl-shaped potential function 
\begin{equation}
\psi_{\text{connected}}(\norm{\qij}) = \frac{ \kappa \rcom}{\norm{\qij}\left(\rcom-\norm{\qij}\right)},
\end{equation}
where $\kappa > 0$ is a tuning factor to adjust the bell shape (see left of $\rcom$ in Fig. \ref{fig:potFunc}). Let the safety controllers for $i$ be $\Dq{i}\psi_{\text{connected}}(\norm{\qij})$ and for $j$ $\Dq{j}\psi_{\text{connected}}(\norm{\q_{ji})} = - \Dq{i}\psi_{\text{connected}}(\norm{\qij})$. When those two agents are getting too close $\norm{\qij} \to 0$, the potential $\psi(\norm{\qij})$ goes to infinity, so the gradient controllers are a strong repulsive force that pushes them away from each other. Conversely, when the two connected agents are at risk of losing their communication link $\norm{\qij} \to \rcom$, then $\psi(\norm{\qij}) \to \infty$, which means the gradient-controllers result in a strong, attractive force that brings them closer again. For multiple agents, the control becomes the sum of gradient potential terms of the other agents, and the magnitude of the gradients helps prioritize the critical inter-agent distances $\qij$.

When the agents are disconnected, which is sometimes unavoidable in underactuated settings where strong flows push them apart, we want them to be able to reconnect. 
Given the assumption of an emergency communication backup outlined in Section \ref{sec:ProblemStatement},  we incorporate a second term to restore connectivity among disconnected agents. To the best of our knowledge, this concept was introduced in \cite{yanFlockingMultiagentSystems2021}. While the augmented potential function in \cite{yanFlockingMultiagentSystems2021} uses a square function of distance to $\rcom$ for disconnected agents, we implement the second term in Eq. (\ref{eq:potFuncFlock}) as a square-root function. This is a design choice in the context of \emph{underactuated} agents in a dynamic oceanic environment, where remote flock members can experience strong divergent flows and direct connectivity may be infeasible or undesirable to achieve.
Thus, our approach yields a relatively low attraction force for agents beyond their communication range.

This results in our final potential function $\psi(z): \R_{\geq 0} \to \R_{>0}$ that is also visualized in Fig. \ref{fig:potFunc}:

\begin{equation}
    \begin{aligned}
            &\psi\left(\norm{\qij}\right)=\sigma_{ij}\underbrace{\frac{ \kappa R_{com}}{\norm{\qij}\left(R_{com}-\norm{\qij}\right)} }_{\text{for connected agents}}\\
            & \quad +(1-\sigma_{ij}) \underbrace{\sqrt{\left(\norm{\qij}-R_{com}  + \epsilon \right)}}_{\text{for disconnected agents}}.
    \end{aligned}
    \label{eq:potFuncFlock}
\end{equation}
where $\sigma_{ij}$ is an edge indicator similar to $a_{ij}$ in Section \ref{subsec:graph_theory}, but with a switching threshold $\epsilon > 0$ inducing a hysteresis when adding new edges, see Eq. (\ref{eq:indicatorFunc}). Hence, $\psi\left(\norm{\qij}\right)$ switches between two terms whether the pair of agents $(i,j)$ are within communication range ($\sigma_{ij}=1$) or disconnected ($\sigma_{ij}=0$). Following the notation of \cite{yanFlockingMultiagentSystems2021}, we define:
\begin{equation}
\begin{aligned}
\sigma_{ij}[t] 
= \begin{cases}0, & \text { if }\left(\left(\sigma_{ij}\left[t^{-}\right]=0\right)
\cap\left(\left\|\qij\right\| \geq R_{com}-\varepsilon\right)\right) \\
& \cup\left(\left(\sigma_{ij}\left[t^{-}\right]=1\right) \cap\left(\left\|\qij\right\| \geq R_{com}\right)\right), \\
1, & \text { if }\left(\left(\sigma_{ij}\left[t^{-}\right]=1\right) \cap\left(\left\|\qij\right\|<R_{com}\right)\right) \\
& \cup\left(\left(\sigma_{ij}\left[t^{-}\right]=0\right) \cap\left(\left\|\qij\right\|<R_{com}-\varepsilon\right)\right),\end{cases}
\end{aligned}
\label{eq:indicatorFunc}
\end{equation}
This hysteresis mechanism avoids constant switching of the dynamical network with multiple agents for edges close to $\rcom$ and helps preserve connectivity in reactive control schemes \cite{jiDistributedCoordinationControl2007}.

The final safe interaction controller for each agent $i$ with maximum propulsion $U_{max,i}$ is then defined as
\begin{equation}
    \pisafei(\q)= - \frac{\sum_{j=1}^{N} \Dq{i}\psi\left(\norm{\qij}\right)}{\norm{\sum_{j=1}^{N} \Dq{i}\psi\left(\norm{\qij}\right) }} U_{max,i}
    \label{eq:uisafe}
\end{equation}

\subsection{Low Interference Safe Interaction Controller}
\label{subsec:lisic}
For our $\pilisic$ that trades off the performance inputs $\uperf$ with the safety input $\usafe$, we propose an approach that weights these control vector inputs for each agent $i$ depending on the risk of losing connectivity or colliding.

\begin{equation}
    \small
    \nonumber
    \ui = \pilisici(\uperf, \usafe) =c_{i}^{(1)} \uisafe  + c_{i}^{(2)} \uiperf, \; \forall i \in \V
    \label{eq:c1}
\end{equation}
where $c_i^{(1)}$ and $c_i^{(2)}$ are weighting factors. Note that $\uisafe \!=\! \pisafei(\q)$ depends on the other agents' positions to guarantee safe interactions.

When collisions or connectivity losses are imminent, $\ui$ should be able to rapidly tend to $\uisafe$ to prioritize the safe interaction safety over performance, i.e. $c_{i}^{(1)} \to 1$ and $c_{i}^{(2)} \to 0$ (Fig. \ref{fig:head_picture} B, C). Conversely, when the network is well connected and there is low danger of collisions, $\ui$ should align with $\uiperf$ to have low interference with the agent's performance control, i.e. $c_{i}^{(1)} \to 0$ and $c_{i}^{(2)} \to 1$ (Fig. \ref{fig:head_picture} A).

Hence we defined a weighting function $\alpha(\q): \R^N \to [0,1]$ such that $c_i^{(1)} = \alpha(\q)$ and $c_i^{(2)} = 1-\alpha(\q)$, see an example in Section \ref{sec:simulation}.
This function $\alpha(\q)$ measures the urgency of $\ui$ to converge to $\uisafe$ and we define it 
\begin{equation}
    c_i^{(1)} = \alpha \left(\norm{\sum_{j=1}^{N} \Dq{i}\psi\left(\norm{\qij}\right) } \right)
    \label{eq:weightingAlpha}
\end{equation}
The function $\alpha$ can be thought of as a monotonically increasing safety activation function taking values between $[0,1]$ depending on its argument's (unbounded) magnitude.
From the definition of $\small\psi\left(\norm{\qij}\right)$ in Eq. (\ref{eq:potFuncFlock}), $\small\lim_{\norm{\qij} \to 0} \psi\left(\norm{\qij}\right) = \infty$ and $\small\lim_{\norm{\qij} \to \rcom} \psi\left(\norm{\qij}\right) = \infty$. Hence, in critical situations $\small \norm{\sum_{j=1}^{N} \Dq{i}\psi\left(\norm{\qij}\right)}$ gets very large so that $c_i^{(1)}$ saturates to $1$ and $c_i^{(2)}$ to $0$, thus prioritizing the network safety for the concerned agents $i \in \V$ i.e. $\ui \to \uisafe$, over each agent individual objective $\uiperf$.

\begin{figure}[t]
    \vspace{0pt}
    \includegraphics[width=0.46\textwidth, trim={1.5cm 1cm 1.7cm 1cm}, clip]{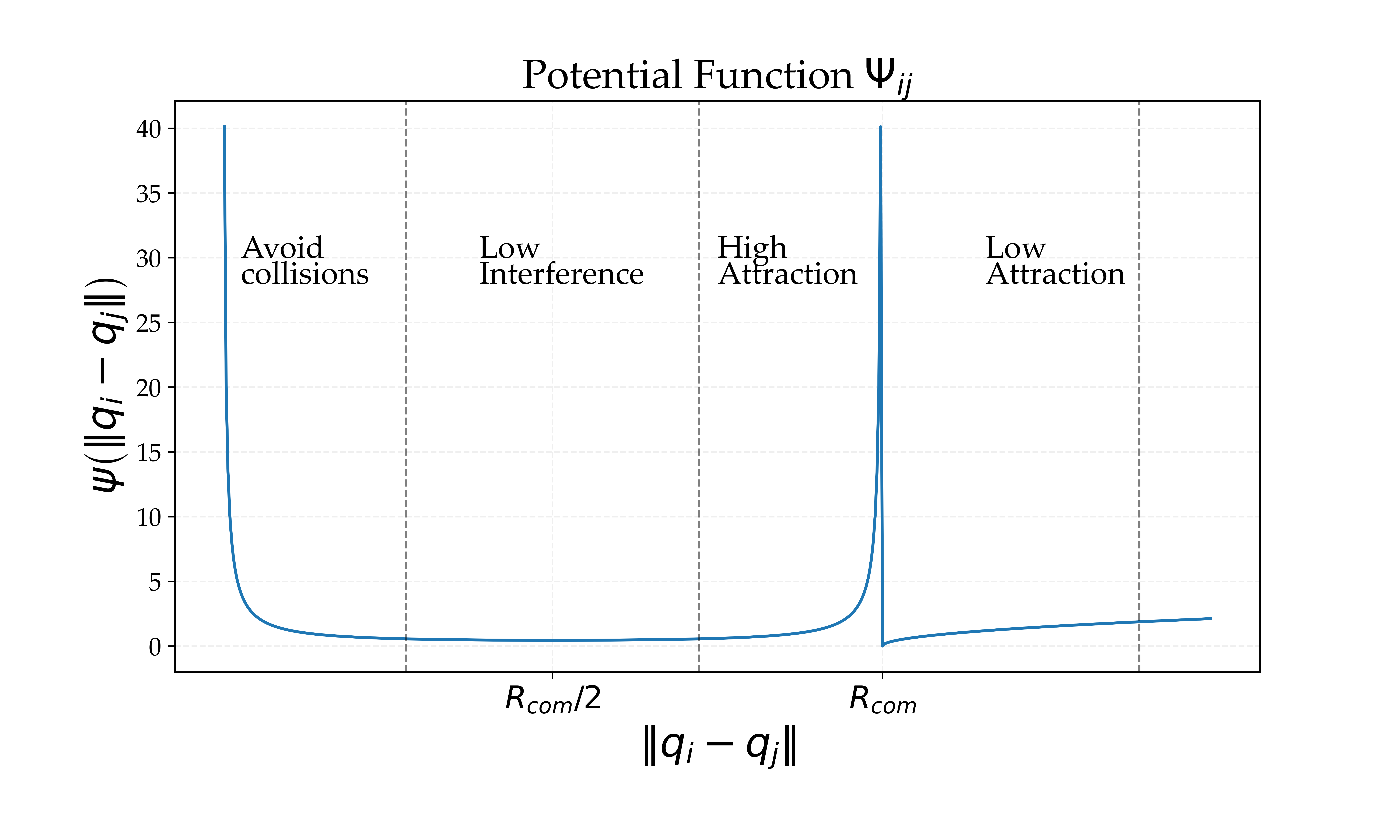}
    \caption{\small Augmented potential function, with two terms to account for agents within and outside the communication range $R_{com}$. A high $\kappa$ parameter is shown to increase the steepness of the slope around $\frac{R_{com}}{2}$, depending on how achieving the exact ideal distance is valued.}
    \label{fig:potFunc}
    \vspace{0pt}
\end{figure}

In other words, $\psi\left(\norm{\qij}\right)$ has a contractivity property for agent inter-distances at the boundaries of the safe set, defined by $0$ and $\rcom$, similarly to \ac{CBFs} \cite{CBF}. With this design, we ensure that agents coming from a disconnected status $\sigma_{ij}\left[t^{-}\right]=0$ to a connected status $\sigma_{ij}\left[t\right]=1$ experience a strong attracting gradient $\uisafe$ to avoid escaping the communication range again. From Fig. \ref{fig:potFunc}, it is also clear that when the network is close to being ideally connected, the gradient norm of the potential function $\norm{\sum_{j=1}^{N} \Dq{i}\psi\left(\norm{\qij}\right)}$ is low so that agent's $i$ control input is dominated by the performance controller since $c_i^{(1)} \to 0$ and $c_i^{(2)} \to 1$.\\

\section{Theoretical Analysis}
\label{sec:theoreticalAnalysis}
 This section analyzes under which conditions our safe interaction controller can maintain connectivity and avoid collisions \cite{nicolas2023multi}. Our analysis follows the common approach to demonstrate that a flock converges to a lattice structure while preventing inter-agent collisions using energy-based analysis and LaSalle's invariance principle \cite{olfati-saberFlockingMultiAgentDynamic2006}. We start by introducing a moving referential frame for the structural collective dynamics \cite{olfati-saberFlockingMultiAgentDynamic2006} with respect to the flock centroid $\q_c$. The relative coordinates are given by $\tqi=\qi-\qc$ and $\tqij=\tqi-\tqj =\qij$. Therefore, $\psi\left( \norm{\qij} \right) = \psi\left( \norm{\tqij} \right)$, and the \emph{total tension energy} or potential energy for the structural dynamics in the relative coordinates yields
\begin{equation}
    H(\tilde{\q}) = \frac{1}{2} \sumi \sumwithout{i}{j} \psi\left(\norm{\tqij}\right)
    \label{eq: totalEnergy}
\end{equation}
A possible approach, although conservative, is to show that a global tension energy decrease of the system $\dot{H} = \sum_{i=1}^n \dot{H}_i \leq 0$ can be achieved by guaranteeing local tension energy decrease $\forall i \in \V$. Assume that $G(t)$ switches at time $t_l$ for $l=0,1,2 \ldots$ and $\dot{H} \leq 0$ on each $[t_l, t_{l+1})$. Then, $H(t_k) = H(t_k^{-}) + m_k \psi \left( \lVert R_{com} - \epsilon \rVert \right)$ \cite{yanFlockingMultiagentSystems2021} with $m_k$ the number of edges added at switching time $k$. The energy can be bound for any subsequent time as the graph topology becomes fixed after a certain time, and only a finite number of maximum edges can be added.
The time-derivative of $H_i$ along the trajectory of agent $i$ yields
\begin{equation}
    \dot{H}_i = \dtq_i^{\top} \sumwithout{i}{j} \Dtq{i}\psi\left(\norm{\tqij}\right)
    \label{eq:Hidot}
\end{equation}
where we exploited the relation $\Dq{i} \psi\left(\norm{\qij}\right) = -\Dq{j}\psi\left(\norm{\qij}\right)$. We seek a condition linking the maximum actuation power of each agent $U_{max,i}$ to the dynamics of the flock, subject to the nonlinear flow $v$. For ease of understanding, assume holonomic actuation i.e. $g(\qi, \ui, t) = \ui$, then $\uisafe \!=\! \pisafei(\q)$ can be directly substituted with
Eq. (\ref{eq:uisafe}). Using $\dtq_i =\dq_{i}-\dq_c$ and the Cauchy-Schwarz inequality in Eq. (\ref{eq:Hidot})  yields:
\begin{equation}
\norm{ c_{i}^{(2)} \uiperf(\qi) + v\left(\qi \right) - \ave{\left(\dq_{Ni} \right)} } \leq c_{i}^{(1)} U_{max,i} \label{eq:HiInequality} \\
\end{equation}
where $\ave{(\cdot)}$ denotes the average and the set $N_i = \mathcal{V} \setminus \{i\}$ the neighboring agents of $i$. More details about this proof can be found in \cite{nicolas2023multi}. A similar inequality to Eq. (\ref{eq:HiInequality}) can be derived for the general dynamics defined in Eq. (\ref{eq:system_dynamics}) if $g(\cdot)$ is a linear map. \\
Let us interpret Eq. (\ref{eq:HiInequality}). The dynamics of the neighboring agents of $i$ depend on their surrounding flows and respective individual control inputs, i.e. $\ave{\left(\dq_{Ni} \right)} = \ave{ \left(v(\q_{Ni})\right)} + \ave{\left(\vect{u}_{Ni} \right)}$.
Despite strong flows, the agents do not necessarily need to be overactuated to meet a local energy decrease $\dot{H}_i \leq 0$. Eq. (\ref{eq:HiInequality}) can be fulfilled even if $\norm{v\left(\qi \right)} > U_{max, i}$, since $v\left(\qi \right)$ can be compensated by $\ave{\left(v(\q_{Ni})\right)}$. In other words, agents in strong flows could still maintain connectivity and avoid collisions as long as the currents experienced by each agent and its neighbors are of similar direction and magnitude. The neighboring flocking control inputs average $\ave{\left(\vect{u}_{Ni} \right)}$ also helps accounting for the current difference term $v\left(\qi \right) - \ave{ \left(v(\q_{Ni})\right)}$. Under these assumptions, we can show that $\dot{H} \leq 0$, which allows to bound the maximum energy and apply LaSalle's Invariance Principle \cite{yanFlockingMultiagentSystems2021}, \cite{wangFlockingMultipleAutonomous2013}, thus ensuring that no collisions or disconnections happen, since $\psi(\norm{\qij}) \to \infty$ when $\norm{\qij} \to 0$ or $\norm{\qij} \to \rcom$. However, for strong divergent flows between agents, it can happen that $\norm{v\left(\qi \right) - \ave{v(\q_{Ni})}} \gg U_{max, i}$ due to the underactuated nature of the agents, which makes satisfaction of Eq. (\ref{eq:HiInequality}) challenging. Note that Eq. (\ref{eq:HiInequality}) is sufficient but not necessary to guarantee $\dot{H} < 0$, as negative local energies can compensate for positive ones.

\section{Simulation Study}
\label{sec:simulation}
The proposed scheme is evaluated on realistic ocean currents, seeking a close reproduction of an innovative \ac{CDR} approach \cite{phykos, killer_CDC2023}. We use \acf{mt-HJR} as a performance single agent controller since it generates a value function yielding the time-optimal control everywhere \cite{wiggert_etal_CDC2022}.

\subsection{Experimental Set-Up}
We study the effectiveness of different controllers in maneuvering a two-dimensional \ac{ASV}, with a thrust angle $\theta$ as a control input. The actuation can be assumed holonomic of fixed thrust magnitude $\norm{u}= 0.1$ m/s, as the vessel can turn to a desired $\theta$ within seconds while our sampling times are 10 minutes. We consider a group of identical $N=30$ \ac{ASV}s with omnidirectional communication capabilities, navigating in strong ocean currents $v(\q, t) \in [0.3, 2]$ m/s. Each single agent's objective is to reach a target region common to all \ac{ASV}s, which could be identified as an ideal seaweed growth region in the context of floating robotic farms \cite{phykos}. Next, we detail the creation of an extensive set of missions to illustrate trade-offs between single-agent objectives and flock connectivity maintenance in a realistic ocean environment.

\begin{figure}
    \centering
    \includegraphics[width=0.44\textwidth, trim={1.5cm 3.5cm 0.1cm 3.25cm}, clip]{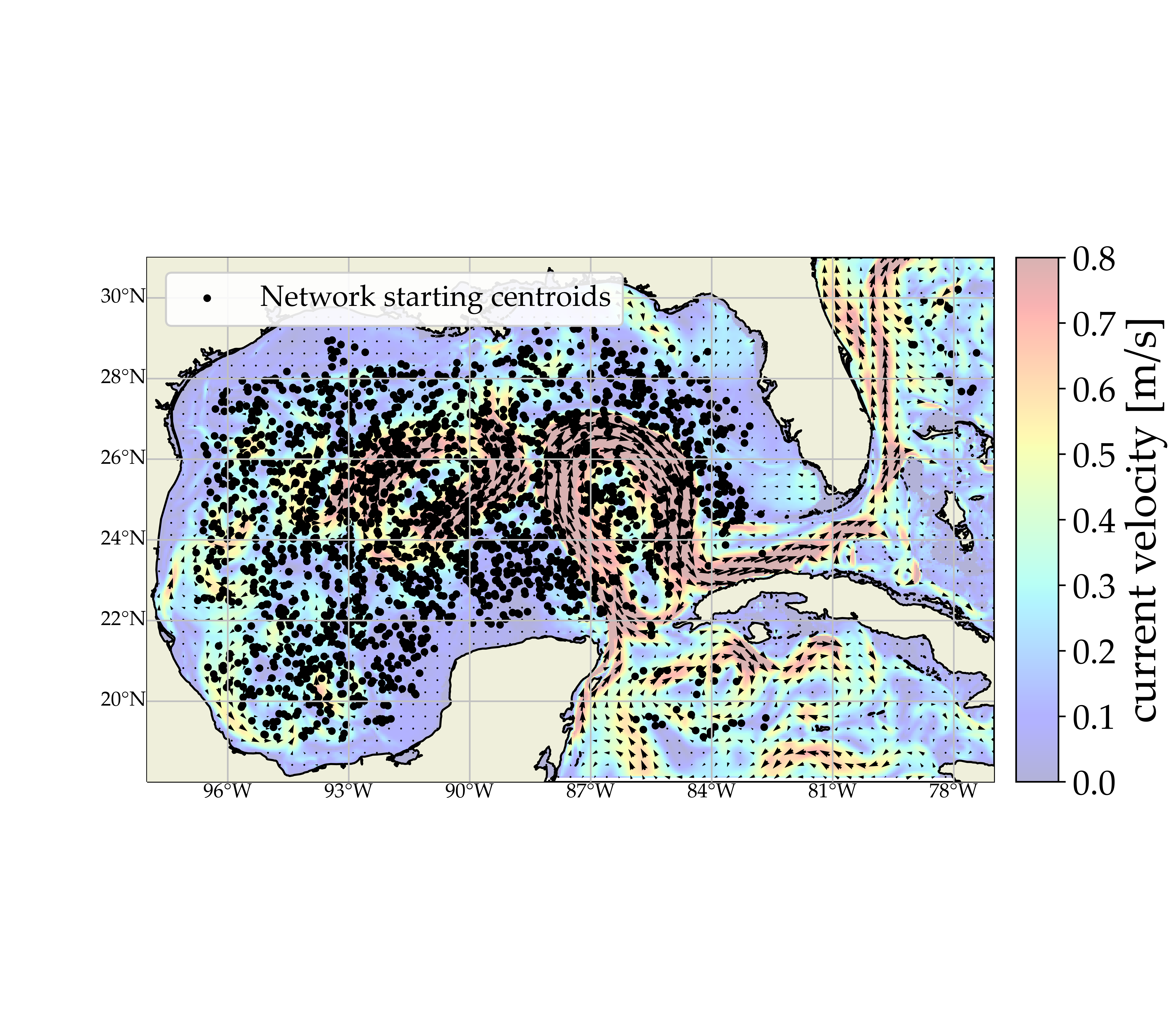}
    \caption{\small 
    We sample a large set of missions $\lvert \M \rvert = 1000$ in the Gulf of Mexico that are spatially and temporally representative of realistic scenarios.}
    \label{fig:sampling}
\end{figure}

\paragraph{Realistic Simulation of Ocean Conditions}
Inspired by \cite{wiggert_etal_CDC2022}, we focus on the Gulf of Mexico region (Fig. \ref{fig:sampling}), as it presents challenging currents. Moreover, we employ two ocean current data sources, which we refer to as HYCOM \emph{hindcasts} \cite{chassignet2009us} and Copernicus hindcasts \cite{CopernicusGlobal} that we use as \emph{forecasts} for realistic scenarios. In our context, the ocean forecast data represents predicted currents $\hat{v}_{FC}$ while the hindcast ocean data are true flows $v$. While the forecast error affects the optimality of the performance \ac{mt-HJR} controllers, the advantage of our reactive safety controller design over predictive schemes \cite{carron2023multi, filotheouDecentralizedControlUncertain2018} is that it is not affected by the error on forecasted currents. We propose two settings to investigate our approach, namely (a) performance \ac{mt-HJR} planning on \emph{hindcasts} and simulation on \emph{hindcasts} (HC-HC) and (b) performance \ac{mt-HJR} planning on \emph{forecasts} and simulation on \emph{hindcasts} (FC-HC). The first allows us to assess performance in an idealized setting where true flows are known, while the second reflects a realistic application in dynamic ocean environments.

\paragraph{Large Representative Set of Missions}
We assume all agents start simultaneously at time $t_0$ a navigation mission to a target region $\T$. The navigation objective of each \ac{ASV} is to reach $\T$ from their start states $\left(\q_1(t_0) \ldots \q_n(t_0) \right )$ within a maximum allowed time $T_{timeout}$. The target $\T$ is defined as a circular region with center coordinates $\q_{\T}$ and fixed radius $r_{\T} = 0.1^{\circ}$ around it. To obtain a diverse set of missions $\mathbb{M}$, the starting times $t_0$ are uniformly sampled between April 2022 and December 2022. $T_{timeout}$ is set to $144$h, and the start points are sampled such that the \ac{ASV}s can reach the target within $[72, 144]h$ to ensure that missions are by definition feasible on true flows and temporally representative enough of realistic scenarios. To prevent stranding side effects, we impose a minimum distance of $111$km between the target area and the land and a minimum distance of $40$km between each \ac{ASV}'s initial position and the land.
We generate $\lvert \mathbb{M} \rvert = 1000$ missions of initially connected and collision-free networks, see Fig. \ref{fig:sampling}.

\subsection{Baseline Controllers}
We build on recent work that proposed a reliable \ac{mt-HJR} controller for underactuated agents utilizing complex flows \cite{wiggert_etal_CDC2022}. This approach directly extends to multiple agents with little extra computation.
The feedback controller for agent $i$ can be obtained from an optimal value function $\mathcal{J}^*$ at time $t$ as $\ui(t)^* = \argmin_{\ui \in \mathbb{U}} g(\qi, \ui, t) \cdot \Dq{i}  \mathcal{J}^*(\qi,t)$. \\

All evaluated controllers use the \ac{mt-HJR} formulation as a single agent performance control. Our baseline scheme, called \acf{mt-HJR-B}, involves each agent only utilizing its time-optimal performance control \ac{mt-HJR} without considering multi-agent interactions.  
This baseline provides a reasonable estimation of the likelihood of collisions and communication losses if each agent were to rely solely on its performance control. In addition, we define a second baseline controller, \acf{mt-HJR-RC} adapted from \cite{pereiraDecentralizedMotionPlanning}. This controller operates in three modes: \textsc{achieveConnectivity}, \textsc{maintainConnectivity}, and \textsc{GoToGoal}, which are selected based on the \ac{ASV}s' relative positions. The \textsc{maintainConnectivity} and \textsc{GoToGoal} modes employ a general navigation function for each agent, which we instantiate to our \ac{mt-HJR} performance controller. This approach is easily integrated with the time-optimal control \ac{mt-HJR}, and the reactive control term can be implemented decentralized.

Finally, we denote our \acf{LISIC} approach from Section \ref{sec:method} as \acf{mt-HJR-LISIC}. The single agent performance controller $\uiperf$ is again \ac{mt-HJR}. The trade-off between each agent's navigational objective and the safe network interaction can be tuned with two parameters. First, the potential function shape (Fig. \ref{fig:potFunc}) can be more or less flat around the ideal distance $\rcom/2$. In this application, we set $\kappa=2$. Furthermore, we now detail our weighting scheme for $c_i^{(1)}$ and $c_i^{(2)}$ via the definition of $\alpha$ in Eq. (\ref{eq:weightingAlpha}) as a \textsc{softmax}-like function 
  \begin{equation}
      c_i^{(1)} = \frac{e^{\norm{\sum_{j=1}^{n} \Dq{i}\psi\left(\norm{\qij}\right)}}}{e^{\norm{\sum_{j=1}^{n} \Dq{i}\psi\left(\norm{\qij}\right) }} + e^{\rho}},  \; \forall i \in \mathcal{V}.
 \end{equation}
 where the parameter $\rho \geq 0$ can be adjusted to achieve faster saturation of the potential function gradient term $\uisafe(\q)$.
 
\subsection{Additional Parameters and Metrics}
The upper connectivity bound $\rcom$ in Eq. (\ref{eq:conn_constraint}) and (\ref{eq:indicatorFunc}) is set to $9$km, which corresponds approximately to radio communication capabilities for \ac{ASV}s. The collision lower threshold from Eq. (\ref{eq:coll_av}) is set to $\rcoll=100$m, providing a practical margin, as one would typically do in a real-world implementation. Moreover, we define $\epsilon= 300$m for the edge hysteresis parameter from Eq. (\ref{eq:indicatorFunc}). We use the Euclidean norm to measure inter-agent distances $d(\qi, \qj)$ and the minimum flock center distance to target $d_{min}(\T)$. The parameters used in the experiments are summarized in Table \ref{tab:parameters}.
\begin{table}[htb!]
\centering
\begin{tabular}{l|l|l}
\toprule
Symbol  & Description                & Value  \\
\midrule
$U_{max}$ & \ac{ASV}s maximum actuation & $0.1$m/s \\
$v$ & Time-varying ocean currents & $[0.3, 2]$m/s \\
$\rcom$  & Upper connectivity bound   & $9000$m \\
$\rcoll$ & Lower collision threshold & $100$m   \\
 $r_{\T}$ &   Radius of circular target region  & $0.1^{\circ}$  \\
$\rho$ & Saturation of potential function & $1$ (no units)\\
$\kappa$ & Shape of potential function & $2$ (no units) \\ 
$\epsilon$ & Hysteresis for adding or removing edges & $300m$ \\
$T_{timeout}$ & Duration of a mission & $144$h
\end{tabular}
\caption{\small
\label{tab:parameters} Relevant Simulation and Controller Parameters.}
\end{table}

\subsection{Numerical Results}
The results over apriori known true currents (HC-HC) and realistic scenario (FC-HC) are presented in Table \ref{tab:results}. Both \ac{mt-HJR-LISIC} and \ac{mt-HJR-RC} exhibit superior performance in terms of connectivity and collision metrics compared to the baseline \ac{mt-HJR-B}. Thus, we conduct statistical testing to compare \ac{mt-HJR-RC} and \ac{mt-HJR-LISIC}. Regarding the disconnection and collision rate, we perform a one-sided two-sample z proportion test for \ac{mt-HJR-LISIC} against \ac{mt-HJR-RC}.

\begin{figure}
\vspace{0pt}
\includegraphics[width=0.5\textwidth, trim={0.1cm 0cm 0.2cm 0cm}, clip]{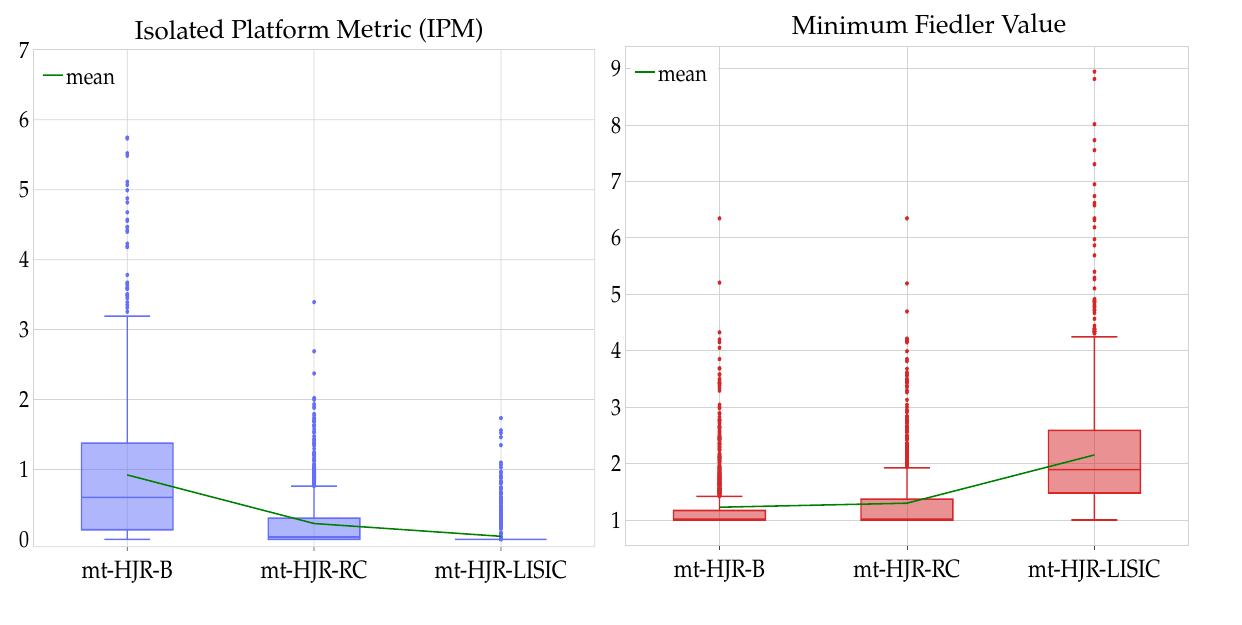}
\caption{\small
Left: \ac{IPM} evaluated on $\mathbb{M}$. Due to its low \ac{IPM}, \ac{mt-HJR-LISIC} typically has both a low disconnection time and a low number of disconnected agents. Right: The minimum Fiedler value $\lambda_2^{min}$ can be used as a graph connectivity measure. A high $\lambda_2^{min}$ ensures better robustness against connectivity failures. }
\label{fig:IPMandFiedler}
\vspace{0pt}
\end{figure}

\begin{figure}
\vspace{0pt}
\includegraphics[width=0.48\textwidth, trim={5.7cm 3.8cm 3.7cm 4cm}, clip]{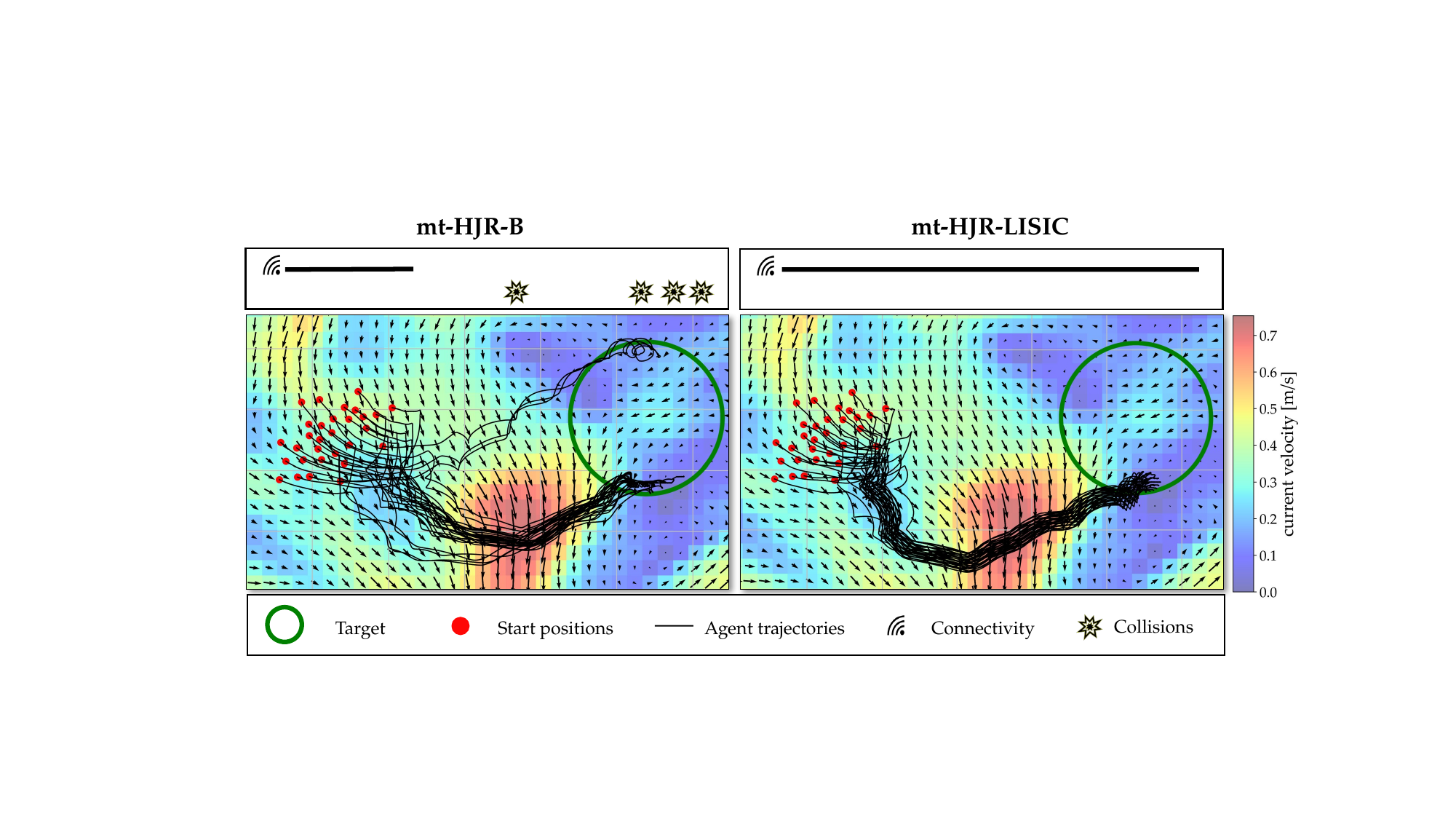}
\caption{\small Mission example with \ac{mt-HJR-B} (left) versus \ac{mt-HJR-LISIC} (right). Note that while the agents' trajectories are depicted for the interval [$t_0$, $T_{timeout}$], the currents in the background represent a snapshot at time $t_0$ and evolve in both direction and magnitude over time. \ac{mt-HJR-LISIC} guarantees communication through the full length of the mission, avoids collisions, and ensures that all agents reach the target.}
\label{fig:flocking}
\vspace{0pt}
\end{figure}

Let $\Gamma$ be the rate collision or disconnection over $\mathbb{M}$ with the null hypothesis $H_0: \Gamma_{\scriptsize \ac{mt-HJR-LISIC}} = \Gamma_{\scriptsize \ac{mt-HJR-RC}}$ to reject in favor of the alternative hypothesis $H_A: \Gamma_{\scriptsize \ac{mt-HJR-LISIC}} < \Gamma_{\scriptsize \ac{mt-HJR-RC}}$. \ac{mt-HJR-LISIC} is statistically significantly better than \ac{mt-HJR-RC} at avoiding disconnections in both (HC-HC) and (FC-HC) scenarios, with p-values of $p=6.3e^{-69}$ and $p=1.7e^{-114}$, respectively. However, it is not significantly better than \ac{mt-HJR-RC} at avoiding collisions. We also perform a Welch's t-test due to the unequal variances of \ac{mt-HJR-RC} and \ac{mt-HJR-LISIC} to test (1) connectivity using the means over $\lvert \M \rvert$ of the \acl{IPM}, i.e., $\mu(\IPM)$ and the minimum Fiedler value recorded over time, i.e., $\mu(\lambda_2^{min})$, (2) performance trade-off with $\mu\left(d_{min}\left(\mathcal{T}\right)\right)$. For both (HC-HC) and (FC-HC) scenarios, \ac{mt-HJR-LISIC} leads to statistically significantly better results for the network connectivity with $p < 1e^{-30}$ for $\mu(\IPM)$ and $\mu(\lambda_2^{min})$ while \ac{mt-HJR-RC} displays a better objective trade-off $\mu\left(d_{min}\left(\mathcal{T}\right)\right)$ with p-values $p < 1e^{-40}$. Moreover, we plot the \ac{IPM} and $\mu(\lambda_2^{min})$, evaluated on the full set of missions $\lvert \M \rvert$ for the three controllers in Fig. \ref{fig:IPMandFiedler}. Among the three evaluated controllers, \ac{mt-HJR-LISIC} has the lowest \ac{IPM}. Because of its higher value of $\mu(\lambda_2^{min})$ (see Fig. \ref{fig:IPMandFiedler}, right), \ac{mt-HJR-LISIC} is more robust against disconnections, and should be the preferred control choice when communication maintenance is prioritized.
Finally, Fig. \ref{fig:flocking} illustrates a navigation mission, comparing a naive multi-agent approach (\ac{mt-HJR-B}) to our safe interaction controller, \ac{mt-HJR-LISIC}.
Note that despite the initial strong currents pushing the \ac{ASV}s away from the desired goal in Fig. \ref{fig:flocking}, the currents eventually shift favorably, allowing the underactuated \ac{ASV}s to reach the target. The \ac{mt-HJR} framework leverages this information through current forecasts to plan intelligently.

\begin{table}[htb!]
\centering
\resizebox{1\columnwidth}{!}{%
\begin{tabular}{llllll}
    \toprule
    & Coll.  & Disconn. & $\mu(\textsc{IPM}) \downarrow$ & $\mu(\lambda_2^{min}) \uparrow$ &  $\mu\left(d_{min}\left(\mathcal{T}\right)\right) \downarrow$ \\ \midrule
    \multicolumn{4}{l}{$\piperf$ plans on true flows: HC-HC} \\ \midrule
    \multicolumn{1}{l}{\ac{mt-HJR-B}} & 68.5\% & 50.1\% & 0.37 & 0.39 &\textbf{ 0} km   \\ 
    \multicolumn{1}{l}{\ac{mt-HJR-RC}} & \textbf{0}\% & 44.8\% & 0.19 &  0.42  &  0.14 km$^*$ \\ 
    \multicolumn{1}{l}{\ac{mt-HJR-LISIC}} & 0.7\% & \textbf{9.9}\%$^*$ & \textbf{0.05}$^*$ & \textbf{1.15}$^*$  & 5.90 km \\  \midrule
    \multicolumn{4}{l}{$\piperf$ plans on forecasts: FC-HC} \\ \midrule
    \multicolumn{1}{l}{\ac{mt-HJR-B}} & 39.1 \% & 70.5\%  & 0.92 & 0.23 &  \textbf{10.55} km  \\ 
    \multicolumn{1}{l}{\ac{mt-HJR-RC}} & \textbf{0}\% & 58\% & 0.23 & 0.30 &10.84 km$^*$ \\ 
    \multicolumn{1}{l}{\ac{mt-HJR-LISIC}} & 0.7\% & \textbf{9.9}\%$^*$  & \textbf{0.043}$^*$  & \textbf{1.15}$^*$  & 13.96 km\\  \bottomrule
\end{tabular}}

\caption{\small
\label{tab:results}We compare the performance of three controllers in two forecast settings. The $^*$ symbol indicates a statistically significant better performance in terms of connectivity, collisions, and distance to the target.}
\end{table}

\subsection{Discussion}
It is clear that \ac{mt-HJR-LISIC} outperforms \ac{mt-HJR-RC} and the \ac{mt-HJR-B} in terms of connectivity metrics. Interestingly, \ac{mt-HJR-LISIC} leads to a slightly higher collision rate in Table \ref{tab:results} than \ac{mt-HJR-RC}. We believe that it is mainly due to two reasons: (1) In \ac{mt-HJR-RC}, the expected risk of collisions is inherently lower as each agent can achieve connectivity with a maximum amount of two other agents \cite{pereiraDecentralizedMotionPlanning} while \ac{mt-HJR-LISIC} achieves a similar structure to a lattice configuration \cite{olfati-saberFlockingMultiAgentDynamic2006} (2) In our example, all agents navigate to the same target, which also increases the risk of collisions, as it is a common implicit regularizer. We expect improvement in collision rate for application to autonomous \ac{ASV}s, where each agent maximizes an objective along its trajectory \cite{killer_CDC2023}. The discrepancy between the performance trade-off with each agent target reaching objective $d_{min}\left(\mathcal{T}\right)$ in Table \ref{tab:results}  is less noticeable in the (FC-HC) setting, since the \ac{mt-HJR} performance is also degraded because of the stochastic error when planning on forecasts \cite{wiggert_etal_CDC2022}.

\section{Conclusion and Future Work}
\label{sec:conclusion}

In this work, we proposed a \ac{H-MAS} approach to maintain network connectivity in complex dynamical flows while satisfying single agent level objectives when feasible. Our method blends a safety controller for collisions and connectivity maintenance with a performance control policy, which allows us to decompose a complex multi-agent problem effectively. 
Our empirical results in realistic ocean dynamics showed that our method efficiently maintains connectivity and avoids collisions in most scenarios while reasonably trading off with each agent's performance objective. Future work involves real-world testing of our experiments, as well as adapting predictive methods \cite{carron2023multi, filotheouDecentralizedControlUncertain2018} to time-varying flows to anticipate disconnections and collisions utilizing the only available coarse ocean forecasts.

\bibliographystyle{IEEEtran}
\bibliography{references,mseas, references_multi_ag}

\end{document}